\let\Xdocument\document
\documentclass{iau}
\let\document\Xdocument

\usepackage{amsmath}
\usepackage{graphicx}
\usepackage{multirow}
\usepackage{hyperref}

\begin{document}

\lefttitle{M. Deka et al.}
\righttitle{A study of the stellar photosphere-hydrogen ionization front interaction in $\delta$ Scuti stars}

\jnlPage{1}{8}
\jnlDoiYr{2023}
\doival{10.1017/xxxxx}
\volno{376}
\pubYr{2023}
\aopheadtitle{Proceedings IAU Symposium}
\editors{Richard de Grijs, Patricia Whitelock and M\'arcio Catelan, eds.}

\title{A study of the stellar photosphere--hydrogen ionization front interaction in $\delta$ Scuti stars}

\author{Mami Deka$^{1}$, Shashi M. Kanbur$^{2}$, Sukanta Deb$^{1,3}$ and Susmita Das$^{4}$}
\affiliation{$^{1}$Department of Physics, Cotton University, Panbazar, Guwahati 781001, Assam, India\\
$^{2}$Department of Physics, State University of New York Oswego, Oswego, NY 13126, USA\\
$^{3}$Space and Astronomy Research Center, Cotton University, Panbazar, Guwahati 781001, Assam, India\\
$^{4}$Konkoly Observatory, Research Centre for Astronomy and Earth Sciences, E\"otv\"os Lor\'and Research Network (ELKH), Konkoly Thege Mikl\'os \'ut 15-17, H-1121, Budapest, Hungary\\
\email{email:mamideka8@gmail.com}}
\begin{abstract}
Pulsating variable $\delta$ Scuti stars are intermediate-mass stars
with masses in the range of 1--3~$\rm M_{\odot}$ and spectral types
between A2 and F2. They can be found at the intersection of the
Cepheid instability strip with the main sequence. They can be used as
astrophysical laboratories to test theories of stellar evolution and
pulsation. In this contribution, we investigate the observed
period--colour and amplitude--colour (PCAC) relations at
maximum/mean/minimum light of Galactic bulge and Large Magellanic
Cloud $\delta$ Scuti stars for the first time and test the hydrogen
ionization front (HIF)-photosphere interaction theory using the
\textsc{mesa}-\textsc{rsp} code. The PCAC relations, as a function of
pulsation phase, are crucial probes of the structure of the outer
stellar envelope and provide insight into the physics of stellar
pulsation and evolution. The observed behaviour of the $\delta$ Scuti
PCAC relations is consistent with the theory of the interaction
between the HIF and the stellar photosphere.
\end{abstract}

\begin{keywords}
stars, variables, delta Scuti, Galactic bulge, Magellanic Clouds
\end{keywords}

\maketitle

\section{Introduction}
The pulsating variable $\delta$ Scuti stars are intermediate-mass
stars with masses in the range of 1-3~$\rm M_{\odot}$ and spectral
types between A2 and F2. They can be found at the intersection of the
Cepheid instability strip with the main sequence. Their pulsation
periods lie between 0.03 and 0.3~days. They pulsate both in
single-mode and multimode. Single-mode $\delta$ Scuti stars obey a
period--luminosity relation (PLR) that makes them excellent distance
indicators \citep{neme17}, whereas multimode $\delta$ Scuti stars are
benchmark objects for asteroseismology.

The stellar photosphere is considered to be at located at an optical
depth $\tau=\frac{2}{3}$ and the hydrogen ionization front (HIF) is
the region of a star where the majority of hydrogen becomes
ionized. The stellar photosphere and the HIF are not always co-moving
during a pulsation cycle. The relative location of the HIF and the
stellar photosphere is pulsation phase-dependent. The HIF interacts
with the photosphere only at some particular phases where the
photosphere lies at the base of HIF. This has been well-established in
Classical Cepheids, RR Lyrae, BL Herculis (BL Her), and W Virginis (W
Vir) stars \citep{simo93, kanb95, kanb96, kanb04, bhar14, ngeo17,
  das18, das20}.

The correlation between the HIF and the stellar photosphere can be
explained using the following equations:
\begin{enumerate}
\item Saha ionization equation:
\begin{align}
\frac{N_{i+1}}{N_i}=\frac{2Z_{i+1}}{n_{\rm e}Z_{i}}{\frac{2\pi m_{\rm e}kT}{h^{2}}}^\frac{3}{2} \exp{-\frac{\chi_{i}}{kT}},     
\end{align}
where $N_{i+1}$ and $N_{i}$ are the numbers of atoms in the $(i +
1)^{\rm th}$ and $i^{th}$ ionization states, respectively; $Z_{i+1}$
and $Z_{i}$ represent the partition functions in the $(i + 1)^{\rm
  th}$ and $i^{th}$ ionization states, respectively; $e_{e}$ is the
number of electrons; $T$ is the temperature; $m_{e}$ is the mass of
the electron, $\chi_{i}$ represents the ionization energy, $k$ denotes
the Boltzmann constant and $h$ is Planck constant. This equation
relates stellar density and temperature.
\item Period is dependent on the density through the period--mean
  density relation:
\begin{align}
P\propto\sqrt{\frac{1}{\rho}}, 
\label{eq:pdens}
\end{align}
where $P$ is the period and $\rho$ represents the stellar density.
\item Stefan--Boltzmann equation:
\begin{align}
\log{T_{\rm max}}-\log{T_{\rm min}}=\frac{1}{10}(V_{\rm min}-V_{\rm max}),
\end{align}
where $\log{T_{\rm max}}$ and $\log{T_{\rm min}}$ are the photospheric
temperature at maximum and minimum light, respectively. $V_{\rm min}$
and $V_{\rm max}$ are the minimum and maximum magnitude of the light
curve, respectively. If $T_{\rm max}$ is independent or weakly
dependent on the pulsation period, then the changes in amplitude are
related to the temperature at minimum light, which will lead to a
correlation between the $V$-band amplitude and the observed colour at
minimum light. Conversely, if $T_{\rm min}$ is independent or weakly
dependent on the period, then a correlation will exist between the
$V$-band amplitude and the observed colour at maximum light.
\end{enumerate}

The engagement between the HIF and the stellar photosphere is
correlated with the temperature of the photosphere, and the
temperature at which hydrogen ionizes is related to the density. As
the colour is related to the temperature and the period is dependent
on the stellar global parameters through the density
(Eq. \ref{eq:pdens}), the relative location of the HIF and the
photosphere at a particular pulsation phase can explain the
sloped/flat period--colour/amplitude--colour (PCAC) relation at that
corresponding phase.

In the present paper, we have verified the HIF and stellar photosphere
interaction theory of \cite*{simo93} in $\delta$ Scuti stars using
their observed PCAC relation and theoretical models computed using
Modules for Experiments in Stellar Astrophysics -- Radial Stellar
Pulsation (\textsc{mesa-rsp}) \citep{smol08,paxt19}.

The remaining part of this paper is organized as follows.
Section~\ref{sec:data} describes the data and methodology
used. Results are discussed in
Section~\ref{sec:results}. Section~\ref{sec:summary} summarizes the
findings of the present work.

\section{Data and Methodology}
\label{sec:data}
We use the optical $(V,I)$-band light curves of $\delta$ Scuti stars
in the Galactic bulge and the Large Magellanic Cloud (LMC) from
OGLE-IV \citep{sosz21} and OGLE-III \citep{pole10}, respectively. The
Galactic sample was cleaned from foreground and background stars
following the procedure described by \citet{piet15} and
\citet{deka22a}. Outliers from the phased light curves were removed
using the median absolute deviation (MAD) \citet{leys13,deka22b}
criterion,
\begin{align*}
\frac{\left|m-\text{Median}(m)\right|}{\text{MAD}(m)} \ge & 3.0,
\end{align*}
where $m$ represents the observed magnitude and $\text{MAD}$ is the
median absolute deviation.

Then, the light curves were fitted with a Fourier sine series
\citep{deb09}:
\begin{align}
\label{eq:Fourier}
m(t)=&A_{0}+\sum\limits_{i=1}^N A_{i} \sin[\frac{2\pi i}{P}(t_{i}-t_{0})+\phi_{i}],
\end{align}
where $A_{0}$ is the mean magnitude; $P$ and $t_{i}$ represent the
period of a star in days and the times of observations, respectively;
$t_{0}$ is the epoch of maximum light and $N$ is the order of the fit,
which set at 4 and 3 for Galactic and LMC light curves, respectively.

The colour at maximum and minimum light is defined as:
\begin{align}
(V-I)_{\rm max} =& V_{\rm max} - I_{\rm phmax},\\
(V-I)_{\rm min} =& V_{\rm min} - I_{\rm phmin},
\end{align}
where $I_{\rm phmax}$ and $I_{\rm phmin}$ correspond to the $I$-band
magnitudes at the same phase as for $V_{\rm max}$ and $V_{\rm min}$,
respectively.

\begin{figure*}
\vspace{0.014\linewidth}
\begin{tabular}{c}
\vspace{+0.01\linewidth}
  \resizebox{1.0\linewidth}{!}{\includegraphics*{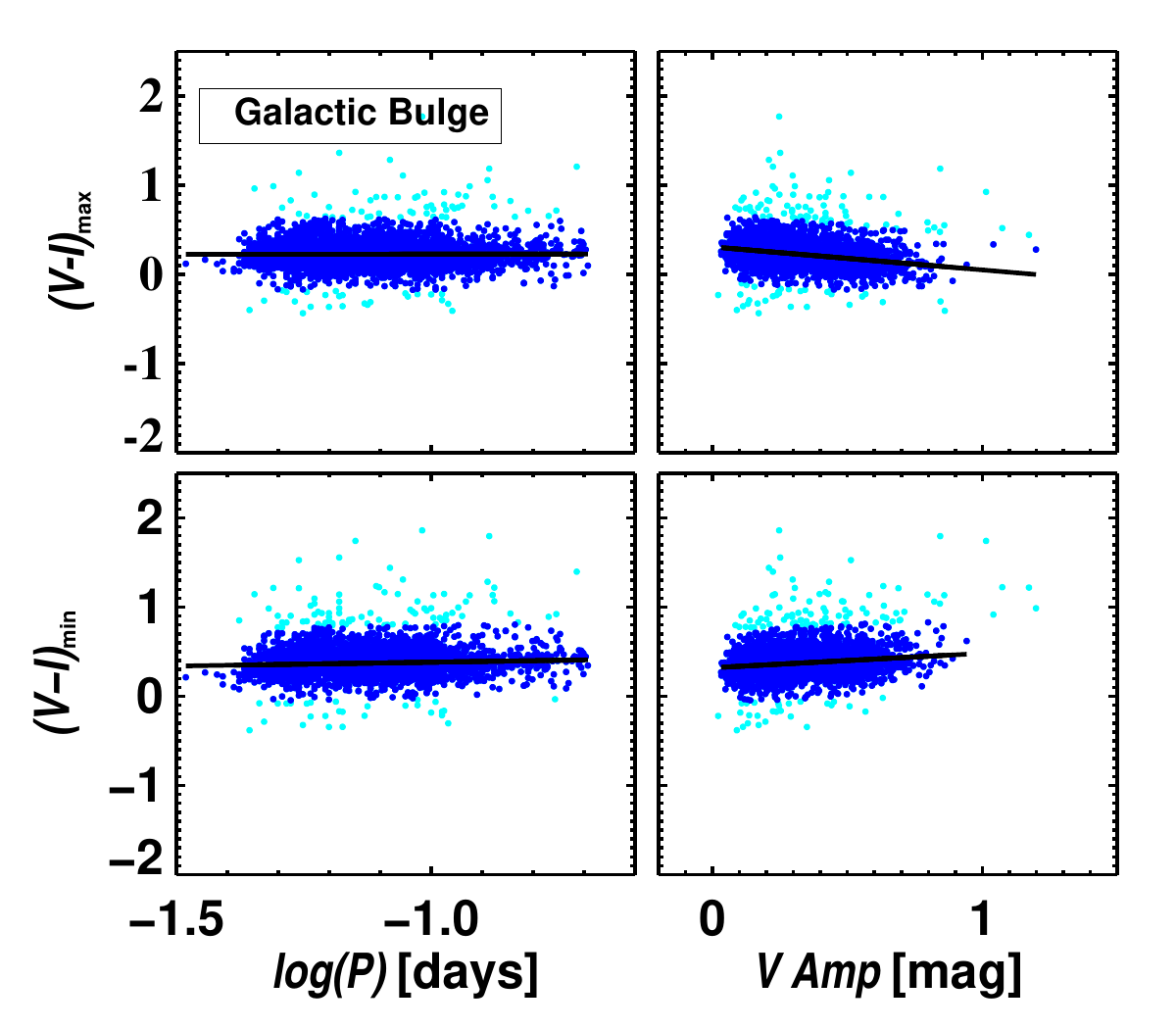}}\\
\vspace{-0.00\linewidth}
\end{tabular}
\caption{PCAC diagram for Galactic bulge $\delta$~Scuti stars. The
  left/right sides are the PC/AC diagrams at maximum (top) and minimum
  (bottom). The cyan points represent the removed outliers. The solid
  lines denote the best fits. }
\label{fig:pcac_blg}
\end{figure*}

\begin{figure*}
\vspace{0.014\linewidth}
\begin{tabular}{c}
\vspace{+0.01\linewidth}
\resizebox{1.0\linewidth}{!}{\includegraphics*{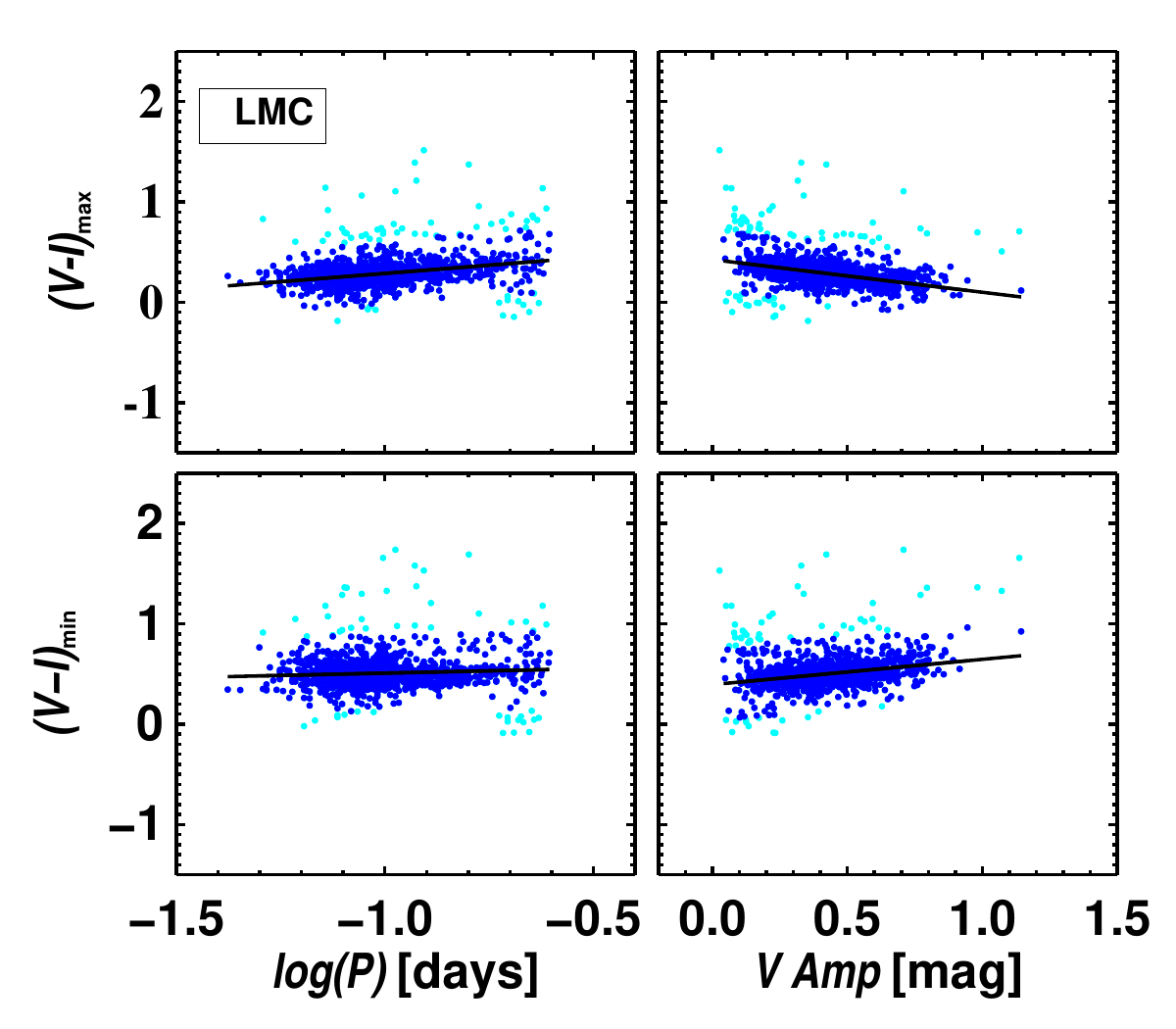}}\\
\vspace{-0.00\linewidth}
\end{tabular}
\caption{As Figure~\ref{fig:pcac_blg} but for the LMC.}
\label{fig:pcac_lmc}
\end{figure*}

The PCAC relations for the $\delta$ Scuti stars were obtained after
correcting the magnitudes and colours for extinction. We used the
reddening maps of \citet{gonz12} and \citet{hasc11} for the Galactic
bulge and LMC, respectively. The PCAC relations were considered after
employing recursive 3$\sigma$ outlier removal.

\begin{table*}
	\centering
	\caption{Slopes and intercepts of the PCAC relation for
          Galactic bulge and LMC $\delta$~Scuti stars. Here, $\sigma$
          denotes the dispersion in the relations.}
    \label{table:result_sa}
	\begin{tabular}{lcccccccr} 
		\hline 
		& & Phase & Slope & Intercept & $\sigma$&Nature of slope\\
		\hline
		
		Galactic bulge&PC & Max & $0.012\pm0.017$ &$0.239\pm0.019$&0.131&Flat\\
		    & & Min & $0.089\pm0.017$ &$0.472\pm0.020$&0.132&Flat\\
		& AC & Max &$-0.252\pm0.013$ & $0.305\pm0.004$ &0.122& Sloped\\
		  &  & Min &$0.166\pm0.014$  &$0.318\pm0.005$&0.131&Flat \\ \hline 
		   
	LMC &	PC & Max & $0.318\pm0.024$ &$0.612\pm0.024$&0.108&Sloped\\
		 &  & Min & $0.117\pm0.027$ &$0.630\pm0.027$&0.124&Sloped\\
	&	AC & Max &$-0.317\pm0.018$ &$0.427\pm0.008$&0.100&Sloped\\
		&   & Min &$0.052\pm0.021$  &$0.488\pm0.009$&0.124&Flat\\ \hline
    \end{tabular}
\end{table*}

\section{Results and Discussion}
\label{sec:results}
\subsection{Observational results}
The PCAC relations for both the Galactic bulge and the LMC are shown
in Figure~\ref{fig:pcac_blg} and Figure~\ref{fig:pcac_lmc},
respectively. The results are summarized in
Table~\ref{table:result_sa}. The PC relations of the Galactic sample
are shallow at both maximum and minimum light. However, for the LMC,
they are sloped at maximum light and shallow at minimum light. The PC
relations for the LMC and Galactic samples at maximum light show
contrasting behaviour. The AC relations exhibit similar behaviour in
both the Galactic and LMC samples.

\begin{figure*}
\vspace{0.014\linewidth}
\begin{tabular}{cc}
\vspace{+0.01\linewidth}
  \resizebox{0.5\linewidth}{!}{\includegraphics*{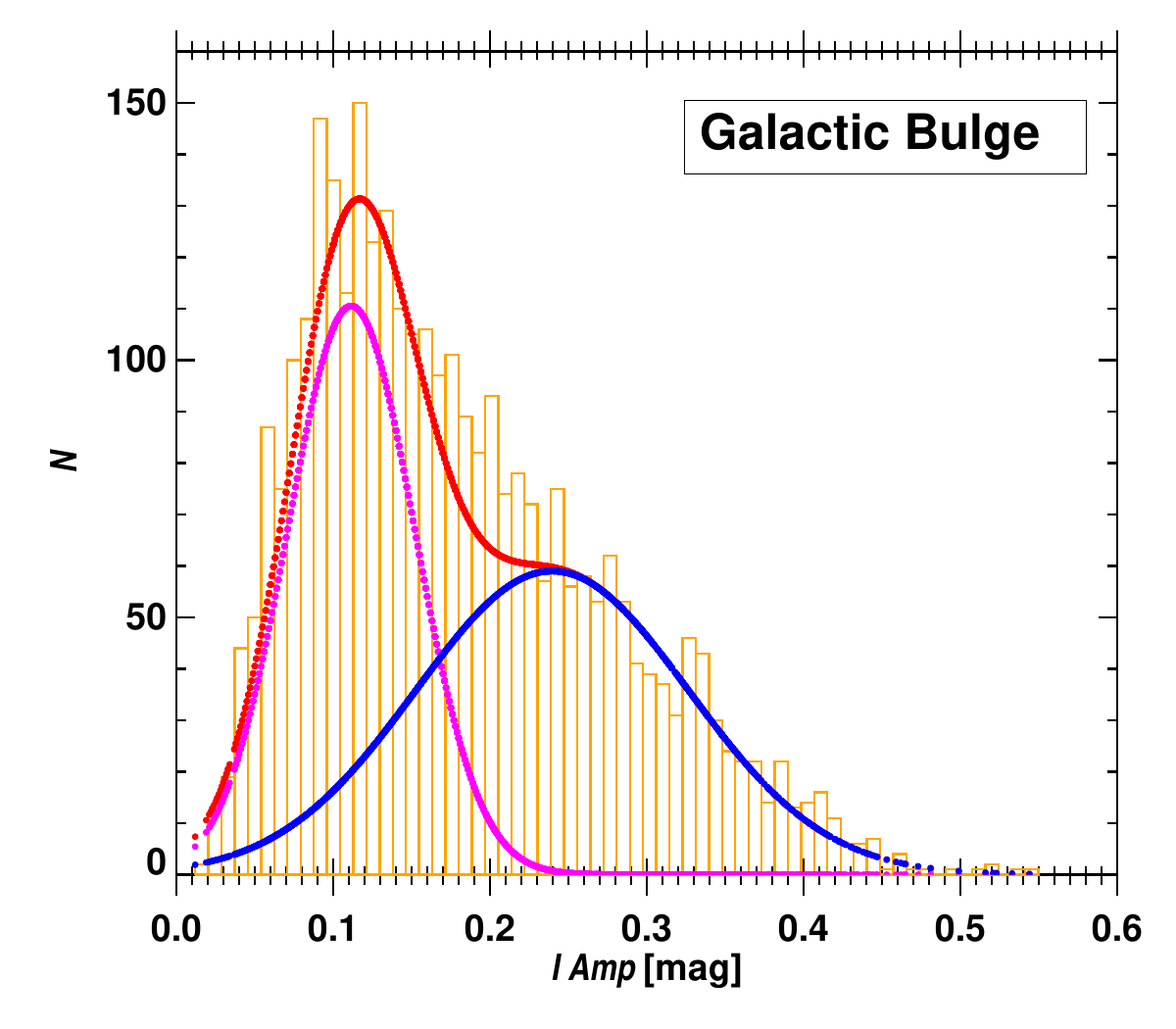}}&
  \resizebox{0.5\linewidth}{!}{\includegraphics*{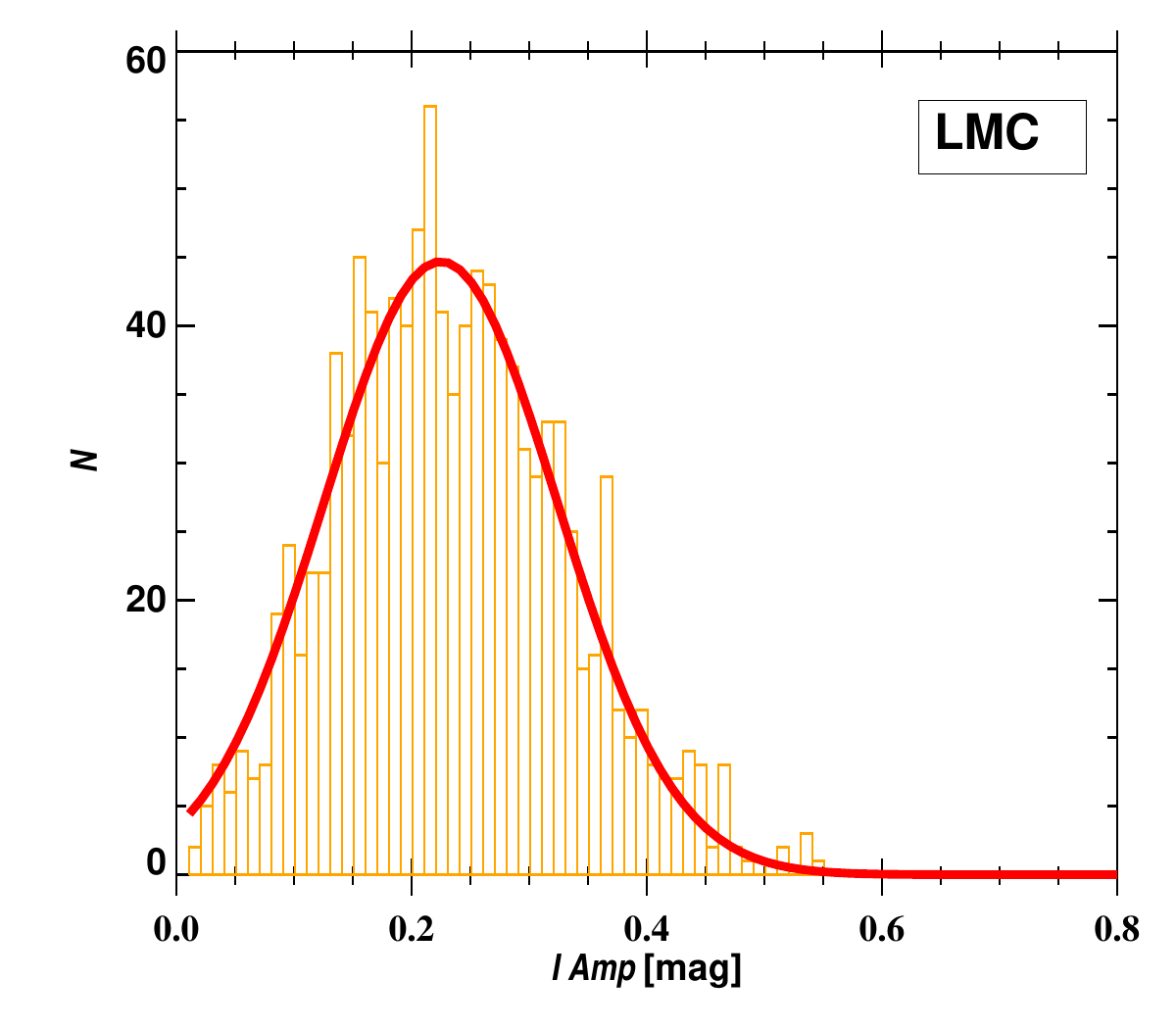}}\\
\vspace{-0.00\linewidth}
\end{tabular}
        \caption{(left) Amplitude histograms in the Galactic
          $I$-band. The red line represents a two-component Gaussian
          fit to both amplitude distributions. The individual Gaussian
          fits are given by magenta and blue lines,
          respectively. (right) LMC amplitude histograms in the $I$
          band. The histogram can be explained with a single Gaussian
          fit (red).}
        \label{fig:gauss_amp}
\end{figure*}

To explain this contrasting behaviour of the PC relations, we have
further investigated the amplitude distributions of both samples,
since amplitude fluctuations are predominantly determined by
temperature fluctuations. A two-component Gaussian function fits the
amplitude distribution of Galactic bulge $\delta$ Scuti stars very
well, as shown in the left panel of Figure~\ref{fig:gauss_amp}.
Separating the two Galactic distributions is beyond the scope of the
present study, and we plan to carry out a more extensive analysis in a
future work. Meanwhile, the LMC distribution is consistent with a
single Gaussian fit, as shown in the right panel of
Figure~\ref{fig:gauss_amp}. Hence, the presence/absence of
low-amplitude stars in the Galactic/LMC samples might have led to the
contrasting behaviour of the PC relations at maximum light.

\subsection{Theoretical results}

\begin{figure*}
\vspace{0.014\linewidth}
\begin{tabular}{cc}
\vspace{+0.01\linewidth}
  \resizebox{0.5\linewidth}{!}{\includegraphics*{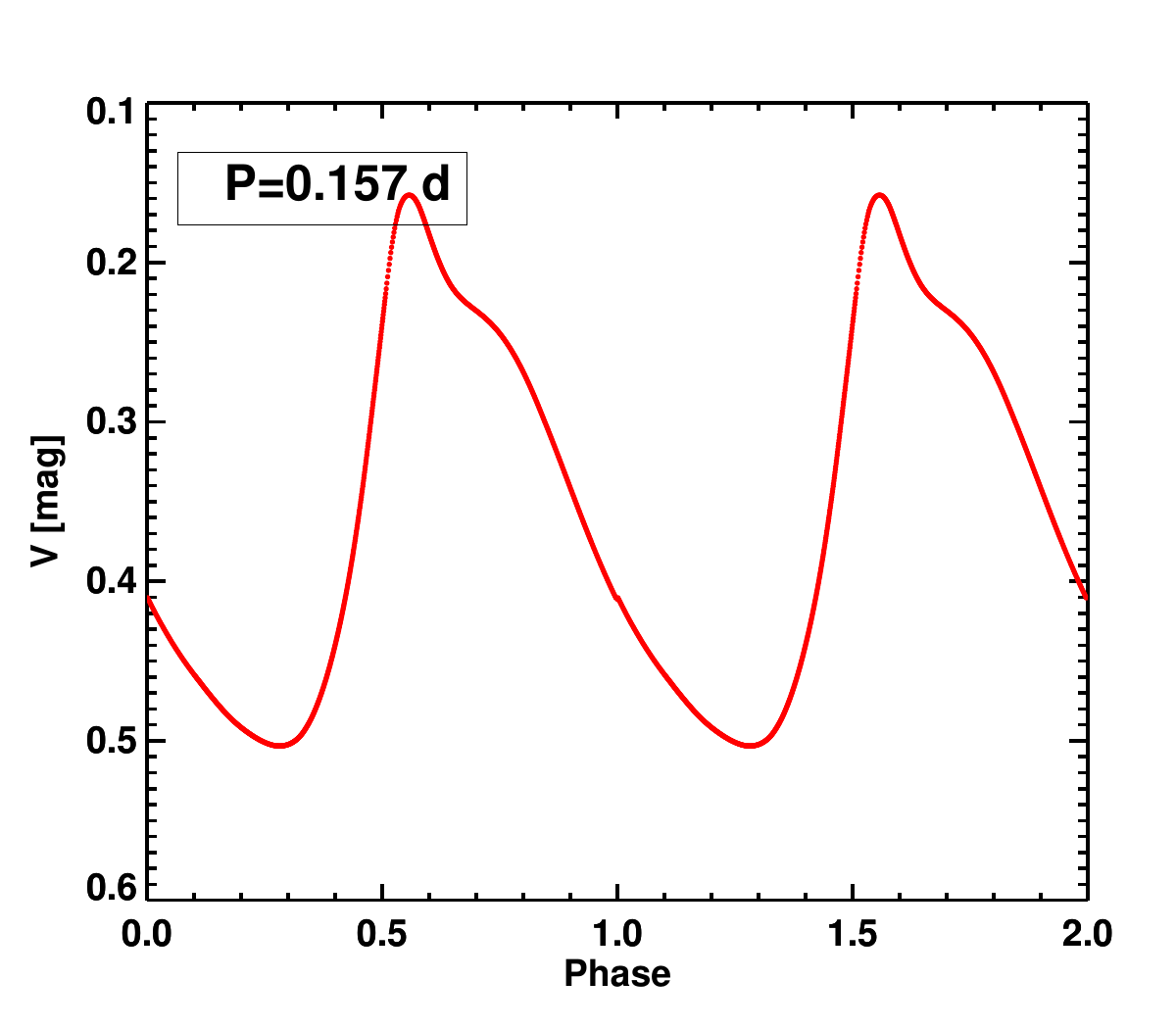}}&
  \resizebox{0.5\linewidth}{!}{\includegraphics*{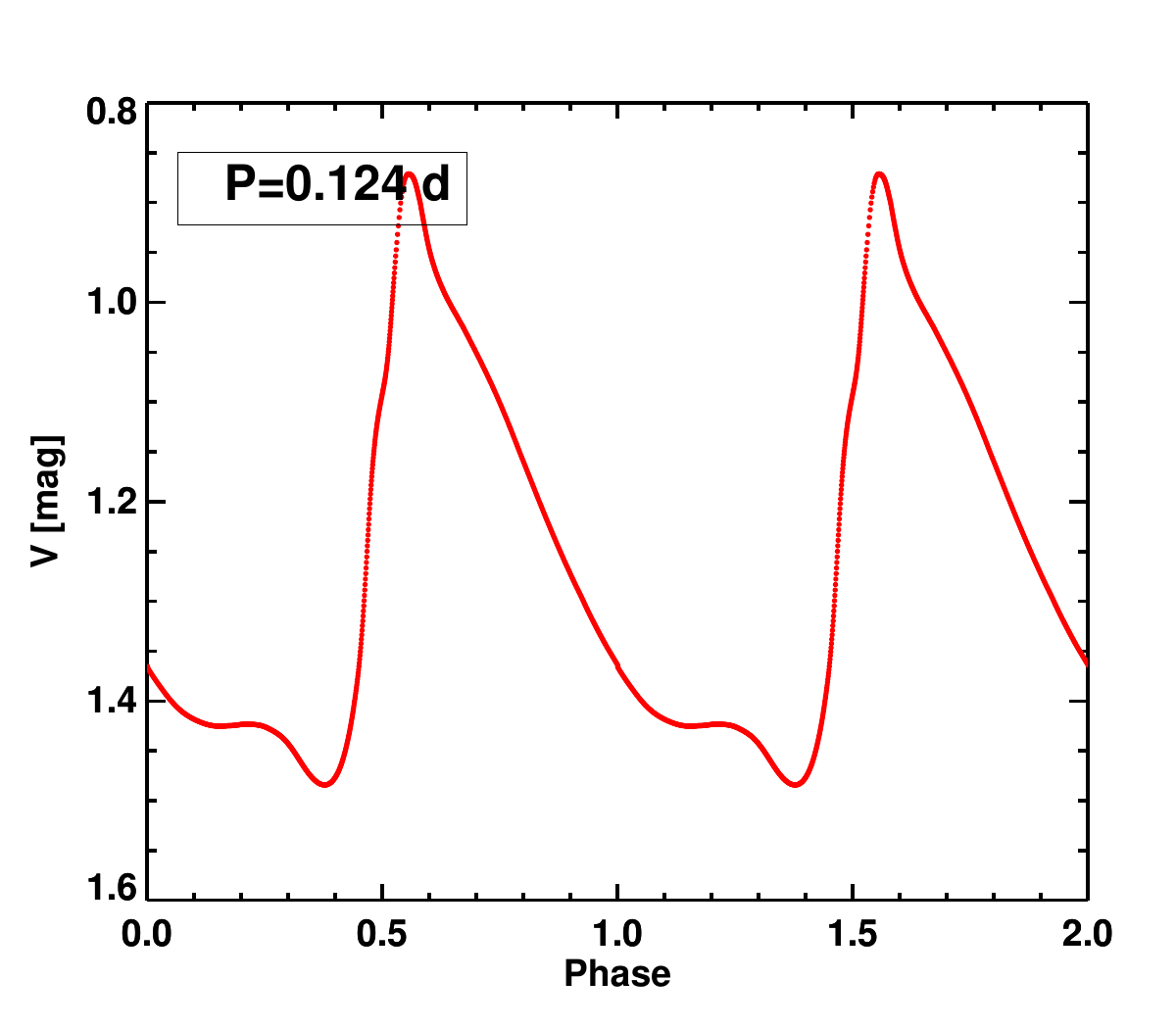}}\\
\vspace{-0.00\linewidth}
\end{tabular}
        \caption{Theoretical light curves of $\delta$ Scuti stars
          obtained using \textsc{mesa}-\textsc{rsp} for (left) a
          Galactic composition and (right) an LMC-like composition.}
        \label{fig:lc}
\end{figure*}

\begin{figure*}
\vspace{0.014\linewidth}
\begin{tabular}{cc}
\vspace{+0.01\linewidth}
  \resizebox{0.5\linewidth}{!}{\includegraphics*{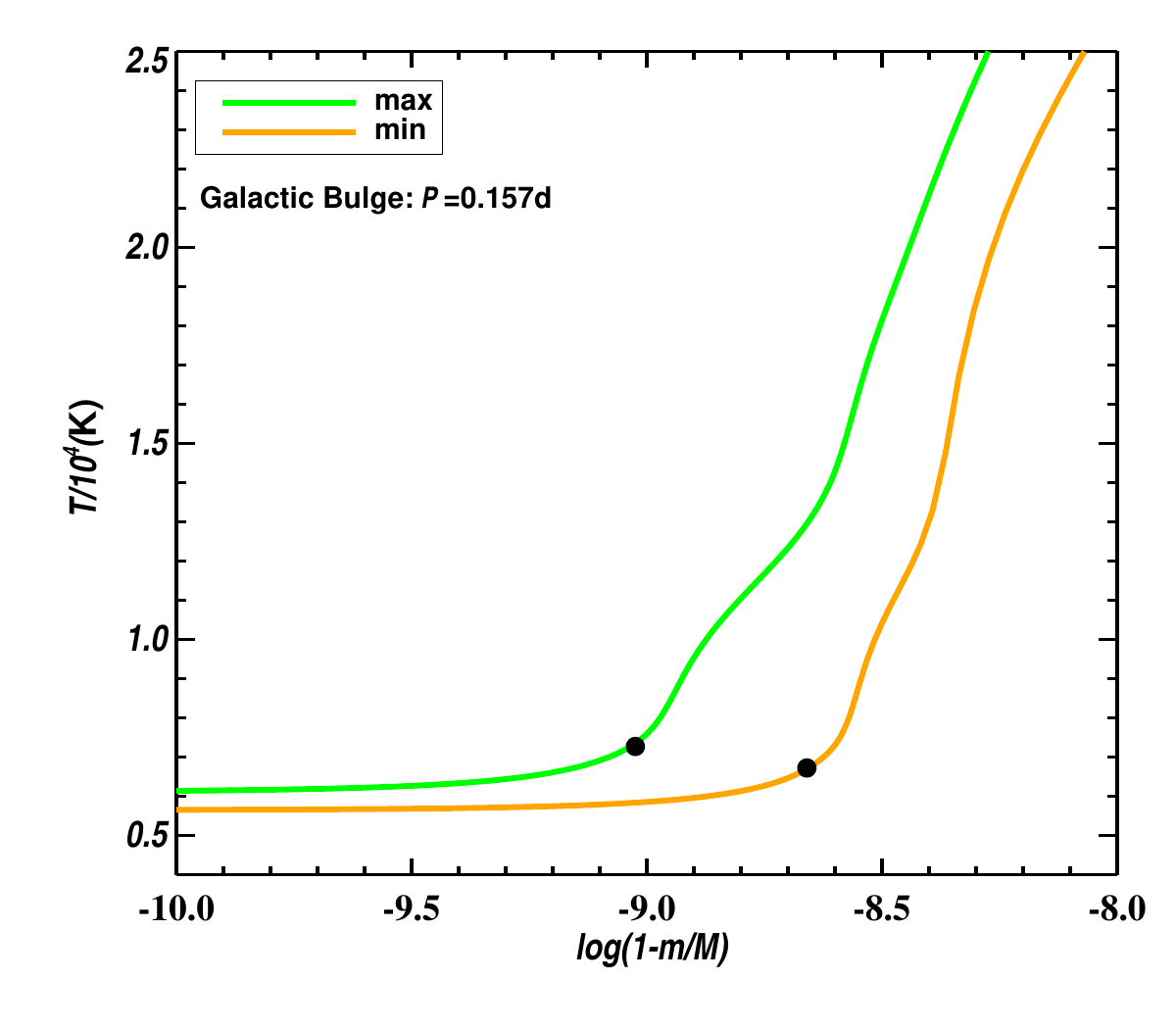}}&
  \resizebox{0.5\linewidth}{!}{\includegraphics*{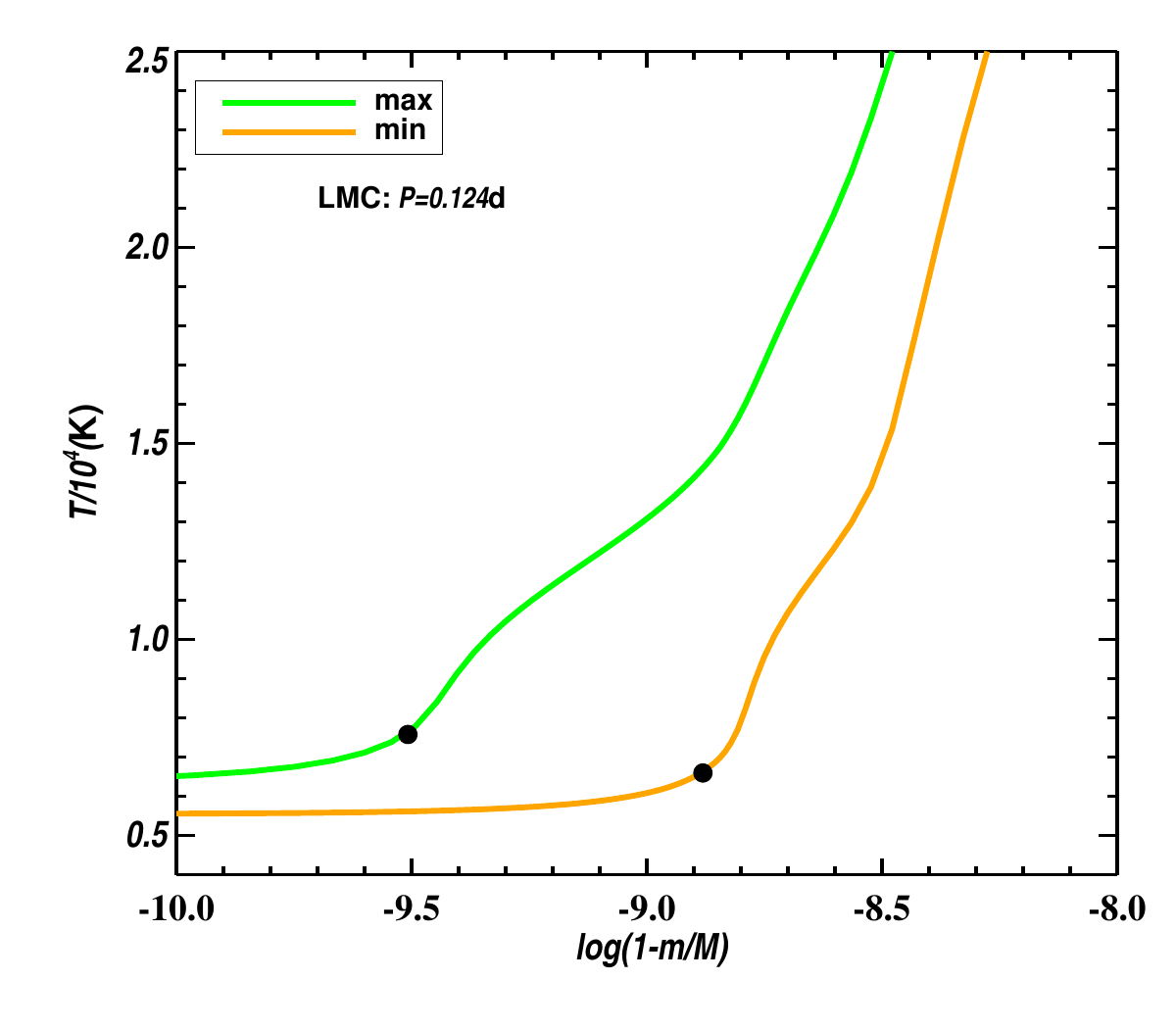}}\\
\vspace{-0.00\linewidth}
\end{tabular}
        \caption{Temperature profile for (left) the Galactic model and
          (right) the LMC model.}
        \label{fig:profile}
\end{figure*}

We have obtained the temperature profiles at maximum/minimum light for
the two models of $\delta$ Scuti stars using \textsc{mesa-rsp},
version `mesa-r15140' \citep{paxt10,paxt13,paxt15,paxt18,paxt19}. The
input parameters taken for the two models were as follows: $Z=0.02,
X=0.70, M=2.0$ M$_{\odot}, L=55$ L$_{\odot}$ and $T=6950$~K (for the
Galactic bulge) and $Z=0.008, X=0.736, M=1.6$ M$_{\odot}, L=25$
L$_{\odot}$ and $T=6900$~K (for the LMC). The light curves obtained
from both models are displayed in Figure~\ref{fig:lc}. The HIF and
photosphere are always found to be engaged for both compositions, as
shown in Figure~\ref{fig:profile}. However, they are engaged at a
lower temperature at minimum light as compared to maximum light, which
explains the smaller observed PC slope. For the LMC model, they are
engaged at a temperature of $\sim7576$~K and for the Galactic model at
$\sim7269$~K at maximum light. This suggests that the HIF is driven
further out in the mass distribution at maximum light for the LMC
model. The temperature fluctuations of Galactic $\delta$ Scuti stars
are smaller than those of their LMC counterparts, leading to smaller
amplitudes. Differences in the locations of the instability strip for
Galactic and LMC stars might have led to the differences in their
amplitudes. The flatter PC relation at maximum light for the Galactic
component is due to the presence of the smaller amplitudes caused by
smaller fluctuations in their photospheric temperature. Hence, the HIF
stellar photosphere theory as described by \citet[][and references
  therein] {simo93,das20} is also consistent with the behaviour of
$\delta$ Scuti stars.

\section{Summary and Conclusions}
\label{sec:summary}
In the present contribution, the $V$- and $I$-band light curves of
$\delta$ Scuti stars from the OGLE-IV and OGLE-III databases
pertaining to the Galactic bulge and the LMC have been used,
respectively, to investigate the corresponding PCAC relations. The
PCAC relations were obtained after employing iterative 3$\sigma$
outlier clipping. The PC relations for Galactic bulge $\delta$ Scuti
stars at maximum/minimum light are shallow, and the AC relation is
sloped at maximum/minimum light. Meanwhile, the PC relations for LMC
$\delta$ Scuti stars at maximum/minimum light are sloped/flat. The
behaviour of the LMC AC relations is similar to those in the bulge,
but the slopes are relatively larger as compared to the Galactic
ones. The amplitude distributions (Figure~\ref{fig:gauss_amp}) reveal
that the Galactic bulge $\delta$ Scuti sample consists of two
populations (low-amplitude and high-amplitude stars), while the LMC
sample consists of mostly high-amplitude stars.

Furthermore, theoretical temperature profiles corresponding to
Galactic and LMC compositions were also obtained using
\textsc{mesa-rsp} to look into the correlation between the HIF and the
photosphere at maximum and minimum light. We find that the HIF and
photosphere are always engaged at maximum and minimum light, for both
models. The temperature at which they are engaged at minimum light is
somewhat lower as compared to maximum light. This explains the flat
observed PC relation at minimum light. However, the difference in the
temperatures at maximum and minimum light for the Galactic model is
relatively smaller as compared to that for the LMC model, which
explains the contrasting behaviour of the observed PC relations to
some extent. Therefore, the relative locations of the photosphere and
HIF can explain the observed PCAC relations of $\delta$ Scuti stars.

\section*{Acknowledgement}
MD thanks the Council of Scientific and Industrial Research (CSIR),
Government of India, New Delhi, for providing a Junior Research
Fellowship (JRF) through CSIR-NET through the research grant
03(1425)/18/EMR-II and she acknowledges travel support provided by
SERB, Government of India vide file number ITS/2023/000385 to attend
IAU Symposium 376, `At the cross-roads of astrophysics and cosmology:
Period-luminosity relations in the 2020s'. S. Deb thanks the CSIR for
financial support received throught CSIR project
03(1425)/18/EMR-II. SMK acknowledges the support of SUNY Oswego and
Cotton University. S. Das acknowledges the KKP-137523 `SeismoLab'
\'Elvonal grant of the Hungarian Research, Development and Innovation
Office (NKFIH). The authors acknowledge the use of the High
Performance Computing facility Pegasus at IUCAA, Pune, India, and the
software \textsc{mesa} r15140
\citep{paxt10,paxt13,paxt15,paxt18,paxt19}.

\bibliographystyle{mnras}
\bibliography{iauguide}

\end{document}